\documentclass[10pt,superscriptaddress,aps,prl,twocolumn,showpacs,nofootinbib, longbibliography, notitlepage,floatfix]{revtex4-2}
\usepackage{etex}
\usepackage{amsmath,amssymb,amsthm}
\usepackage[colorlinks=true,citecolor=blue,urlcolor=blue]{hyperref}

\usepackage{xr} 
\usepackage{times,txfonts}
\usepackage{braket}
\usepackage{color}
\usepackage{natbib}
\usepackage{amsmath,blkarray}
\usepackage{mathtools}
\usepackage{ulem}
\usepackage{latexsym}
\usepackage{tabularx, booktabs}
\usepackage{graphics,epstopdf}
\usepackage{graphicx}
\usepackage{float}
\usepackage{amsfonts}
\usepackage{tikz}
\usepackage[ruled]{algorithm2e}
\usetikzlibrary{quantikz}
\usepackage{color,soul}

\newcommand{\be}{\begin{equation}}
\newcommand{\ee}{\end{equation}}
\newcommand{\ba}{\begin{eqnarray}}
\newcommand{\ea}{\end{eqnarray}}

\usepackage{multirow}
\usepackage{appendix}
\usepackage{url}
\usepackage{pdfpages} 
\usepackage{pgffor} 
\usepackage{xr} 

\makeatletter
\AtBeginDocument{\let\LS@rot\@undefined}
\makeatother

\def\supplementfilename{supplementary}

\pdfximage{\supplementfilename.pdf}
\def\numbersupplementpages{\the\pdflastximagepages}

\newif\ifarXiv
\arXivtrue 

\begin{document}
\title{Lyapunov Controlled Counterdiabatic Quantum Optimization} 
\author{Pranav Chandarana \href{https://orcid.org/0000-0002-3890-1862}{\includegraphics[scale=0.05]{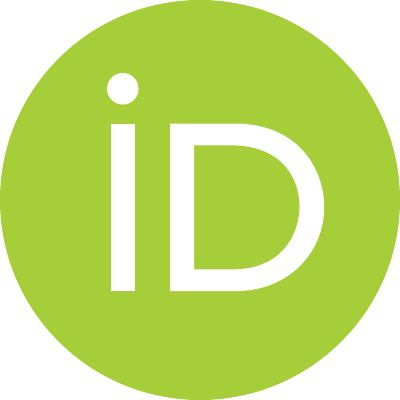}}}
\email{pranav.chandarana@gmail.com}
\affiliation{Department of Physical Chemistry, University of the Basque Country UPV/EHU, 
48080 Bilbao, Spain}
\affiliation{EHU Quantum Center, University of the Basque Country UPV/EHU, 
48940 Leioa, Spain}

\author{Koushik Paul \href{https://orcid.org/0000-0002-2732-9629}{\includegraphics[scale=0.05]{orcidid.pdf}}}
\email{koushikpal09@gmail.com}
\affiliation{Department of Physical Chemistry, University of the Basque Country UPV/EHU, 
48080 Bilbao, Spain}
\affiliation{EHU Quantum Center, University of the Basque Country UPV/EHU, 
48940 Leioa, Spain}

\author{Kasturi Ranjan Swain\href{https://orcid.org/my-orcid?orcid=0009-0008-7227-3856}{\includegraphics[scale=0.05]{orcidid.pdf}}}
\affiliation{Department  of  Physics  and  Materials  Science,  University  of  Luxembourg,  L-1511  Luxembourg,  Luxembourg}

\author{Xi Chen \href{https://orcid.org/0000-0003-4221-4288}{\includegraphics[scale=0.05]{orcidid.pdf}}}
\email{xi.chen@csic.es}
\affiliation{Instituto de Ciencia de Materiales de Madrid (CSIC),
Cantoblanco, E-28049 Madrid, Spain}

\author{Adolfo del Campo\href{https://orcid.org/0000-0003-2219-2851}{\includegraphics[scale=0.05]{orcidid.pdf}}}
\email{adolfo.delcampo@uni.lu}
\affiliation{Department  of  Physics  and  Materials  Science,  University  of  Luxembourg,  L-1511  Luxembourg,  Luxembourg}
\affiliation{Donostia International Physics Center,  E-20018 San Sebasti\'an, Spain}
\affiliation{Theoretical Division, Los Alamos National Laboratory, Los Alamos, NM 87545, USA}

\begin{abstract}
We introduce a quantum algorithm integrating counterdiabatic (CD) protocols with quantum Lyapunov control (QLC) to tackle combinatorial optimization problems. This approach offers versatility, allowing implementation as either a 
digital-analog or purely digital algorithm based on selected control strategies. By examining spin-glass Hamiltonians, we illustrate how the algorithm can explore alternative paths to enhance solution outcomes compared to conventional CD techniques. This method reduces the dependence on extensive higher-order CD terms and classical optimization techniques, rendering it more suitable for existing quantum computing platforms. The combination of digital compression via CD protocols and the adaptable nature of QLC methods positions this approach as a promising candidate for near-term quantum computing.
\end{abstract}

\maketitle

\begin{figure}[t]
    \centering
    \includegraphics[width=0.8\linewidth]{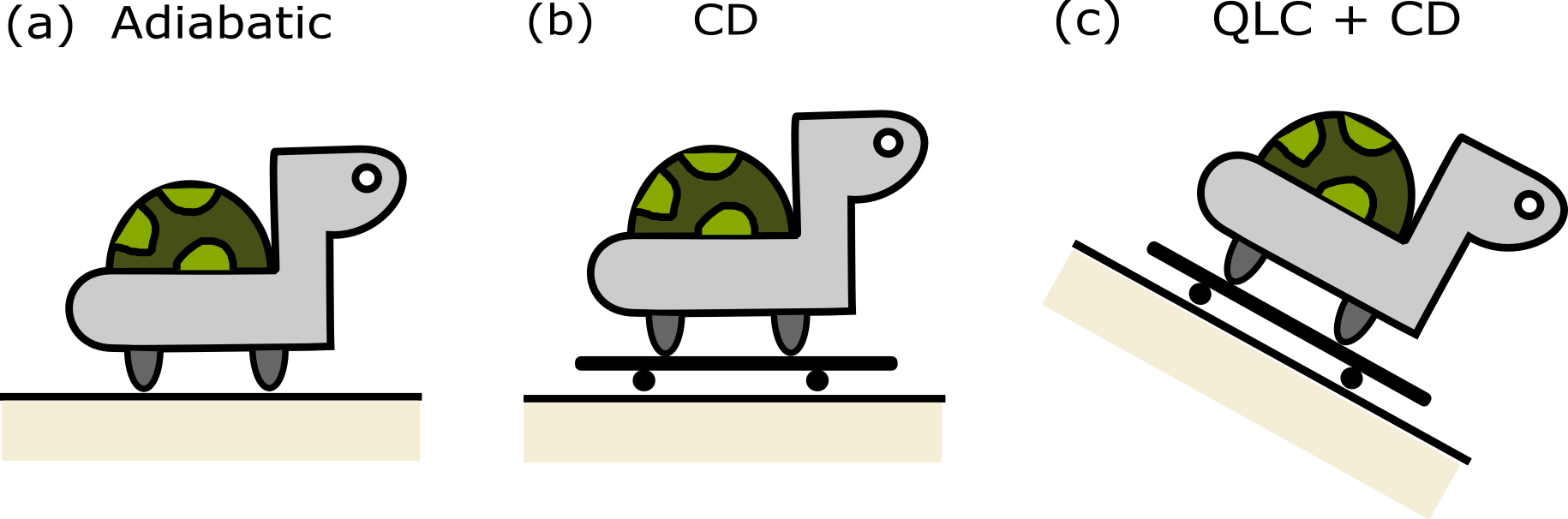}
    \caption{Schematic representation of Lyapunov-controlled CD protocols. (a)  Adiabatic process. 
    (b)  Speed up of CD protocols 
    by the AGP in Eq.~\eqref{eq:impulse}, represented by the skateboard. (c)  The Lyapunov-controlled CD protocol, using the information on the slope of the landscape as in Eq.~\eqref{eq:lyacond}, makes the process even faster.   }
    \label{fig:turtle}
\end{figure}

Adiabatic quantum optimization (AQO) has significantly evolved since its early proposals~\cite{doi:10.1126/science.1057726,doi:10.1142/S021974990800358X} as a paradigm for finding approximate solutions to combinatorial optimization problems. This is accomplished by mapping the problem to a spin-glass Hamiltonian with ground-state as the solution~\cite{10.3389/fphy.2014.00005}. However, showcasing its usefulness is still challenging because AQO requires deep circuits, while near-term quantum devices suffer from limited coherence, restricted connectivity, and noise.
\begin{figure*}[t]
    \centering
    \includegraphics[width=0.95\linewidth]{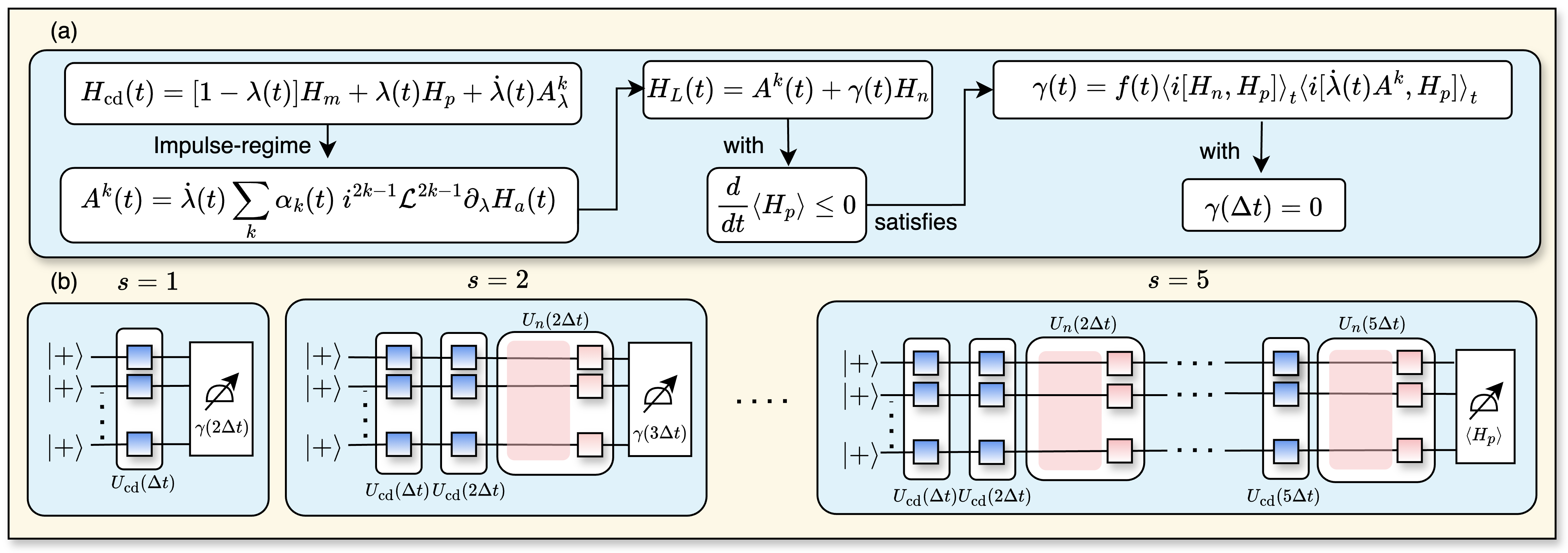}
    \caption{Schematic diagram of the algorithm. Panel (a) illustrates the flow starting from the counterdiabatic Hamiltonian $H_{\rm{cd}}$ and subsequently adding the analog Lyapunov Hamiltonian $\gamma(t) H_n$ and finding $\gamma(t)$ using the Lyapunov condition $\frac{d}{dt}\braket{H_p} \leq 0$. Panel (b) depicts the corresponding digital-analog implementation,  realizing digital CD blocks and utilizing measurements at each step $s$ to find analog QLC times for the next steps. The algorithm computes $\braket{H_p}$ in the last step.  }
    \label{fig:schematicalgo}
\end{figure*}
Digitized counterdiabatic quantum optimization (DCQO)~\cite{PhysRevResearch.4.L042030,PhysRevApplied.15.024038,YaoLinBukov21} has emerged as a paradigm to tackle these challenges by digitally compressing AQO circuits using counterdiabatic (CD) protocols~\cite{Demirplak03,Berry09,delcampo12,Campbell15}.

Consider a Hamiltonian $H_p$, whose ground state encodes the solution to the problem. In DCQO, we construct a modified adiabatic Hamiltonian
\begin{equation}\label{hcd}
    H_{\rm{cd}}(t) = \underbrace{[1-\lambda(t)]H_m + \lambda(t) H_p}_{H_a(t)} + \dot \lambda(t)  A_\lambda^k,
\end{equation}
where, $H_a(t)$ denotes the standard adiabatic Hamiltonian. In AQO, the scheduling function $\lambda(t)$ smoothly drives $H_a(t)$ from an easy-to-prepare ground state of $H_m$ to the ground state of $H_p$. However, a velocity-dependent $k$-order approximate adiabatic gauge potential  (AGP),  $A_\lambda^k = i \sum_k \alpha_k O_{2k-1}$, can be introduced to circumvent the slow-driving condition~\cite{PhysRevLett.123.090602}. Here, $O_k = \mathcal{L}^k \partial_\lambda H_a(t)$, with $\mathcal{L}(\circ) = [H,\circ]$ being the Liouvillian super-operator. The coefficients $\alpha_k(t)$ can be obtained by various methods, including action minimization~\cite{doi:10.1073/pnas.1619826114,PhysRevLett.123.090602} and Krylov subspace methods~\cite{PhysRevX.14.011032,bhattacharjee2023lanczos}. Approximate AGPs are particularly useful for quantum computing since the exact CD terms are highly non-local, and their construction requires the knowledge of the full spectral properties of $H_a(t)$ \cite{delcampo12,Saberi14}. Starting from the ground state of $H_m$, one can construct a $s$-stepped digitized evolution $U(s) \approx \prod_{j=1}^{s} \exp \left\{-i H_{\rm{cd}}(j \Delta t) \Delta t\right\}$ with step size $\Delta t$. The evolution generated by $U(s)$ drives the system faster compared to AQO due to the addition of $A_\lambda^k$ and compresses the total circuit depths significantly~\cite{PhysRevResearch.4.L042030,YaoLinBukov21}.

Recently, it was demonstrated that the circuit depth can be compressed further for a very short time evolution, where $|\alpha_k(t) \dot \lambda(t)|\gg |\lambda(t)|$~\cite{PhysRevApplied.20.014024,cadavid2023efficient}. In this regime, contributions from $H_a(t)$ can be neglected because the CD term governs the system's dynamics. This can be identified as the impulse-regime DCQO with corresponding Hamiltonian 
\begin{equation}\label{eq:impulse}
    A^k(t) = i \dot\lambda(t) \sum_k \alpha_k(t) \mathcal{L}^{2k-1} \partial_\lambda H_a(t).
\end{equation}
Hence, the digitized evolution  is set by $U_{\rm{cd}}(j
\Delta t) =  \exp \left\{-i A^k(j \Delta t) \Delta t\right\}$. Impulse-regime DCQO has shown promising results in both hybrid quantum-classical~\cite{PhysRevApplied.20.014024} and fully quantum setting~\cite{cadavid2023efficient}. However, for the fully quantum setting, this regime is valid only for a short period until the contributions from the neglected terms start to contribute. Meanwhile, hybrid quantum algorithms encounter difficulties with trainability and measurement overhead associated with gradient estimations~\cite{barrenplateau,Wierichs2022generalparameter}.


In parallel, techniques such as the quantum Lyapunov control (QLC) have also been investigated to design optimization algorithms~\cite{PhysRevLett.129.250502}. QLC techniques involve defining a Lyapunov function that will control the dynamics of a given quantum system by implementing a feedback law~\cite{cong2013survey}. These algorithms have been shown to resolve the trainability issues by circumventing the use of classical optimization.
However, due to the feedback law, these methods also require large circuit depths and measurements depending upon the Lyapunov functions implemented.

In this Letter, we propose a quantum algorithm that unifies CD protocols with QLC methods. This algorithm reduces the circuit depth further with enhanced performance by utilizing measurement as a resource and admits
digital and digital-analog implementations, depending on the choice of control Hamiltonians. Recently, proposals for CD-inspired feedback quantum optimization have also been made~\cite{malla2024feedback}, but the control strategy was limited to QLC methods only.
Digital-analog quantum algorithms generally utilize digital quantum computers' flexibility and analog simulations' robustness to neglect the noisy two-qubit gate errors~\cite{PhysRevA.101.022305,PhysRevResearch.6.013280,garcia2021noise}. Here, we achieve the same goal by combining digital CD protocols with analog QLC methods. Specifically, the evolution times of analog blocks are found by utilizing  QLC methods, 
generating new paths to solutions without increasing digital circuit depth.  
On the other hand, the digital algorithm utilizes the feedback to improve the efficiency of higher-order CD terms. Furthermore, both algorithm variants alleviate classical optimization to relax the trainability issues. Henceforth, we develop digital-analog and digital variants and compare them with the state-of-the-art DCQO algorithms to showcase the potential advantage of the proposed approach.


\textit{Digital-analog variant.---} One of the significant advantages of QLC methods is the freedom to choose control Hamiltonians. In contrast, with approximate CD driving the choice is limited to odd-order nested commutators. We exploit this freedom to introduce an additional term $ \gamma(t)H_n$ to Eq.~\eqref{eq:impulse}, where $H_n$ represents the native Hamiltonian of the quantum hardware. This implies that the associated evolution $U_n(j\Delta t) = \text{exp}\{ -i \gamma(j \Delta t)H_n \Delta t\}$ corresponds to an analog block with which the quantum hardware evolves. As we aim for approximate optimization,  we set the Lyapunov function as $\langle H_p\rangle$, where $\langle\circ\rangle$ denotes the expectation value for a given state $\ket{\psi(t)}$. From here, the task  is to find the control parameter $\gamma(t)$ such that 
\begin{equation}\label{eq:lyacond}
    \frac{d}{dt} \braket{H_p}\leq 0.
\end{equation}   

To this end, we propose a digital-analog Lyapunov-controlled counterdiabatic quantum optimization (DALCCO) algorithm, where the Hamiltonian is given by
\begin{equation}\label{eq:dalccoham}
     H_L(t)  =   A^k(t) + \gamma(t)H_n.
\end{equation}
For a specific time $t$, Eq.~\eqref{eq:lyacond} and Eq.~\eqref{eq:dalccoham} can be used to show that if
\begin{equation}\label{eq:DALCCOeq}
    \gamma(t) = f(t) \braket{i[H_n,H_p]}_t \braket{i[  A^k(t), H_p]}_t,
\end{equation}
with $f(t)$ as a Lagrange multiplier, the condition in Eq.~\eqref{eq:lyacond} can be satisfied (see Supplemental Material). For each step $\Delta t$, analog times for the next step can be found by performing measurements. This will constitute our feedback law as required by QLC methods. If the CD term is a local Hamiltonian, the algorithm will feature digital single-qubit gates from the CD protocols followed by analog blocks from QLC methods, resulting in a digital-analog algorithm.

To benchmark DALCCO against DCQO, we select  an $N$-qubit all-to-all connected spin-glass Hamiltonian $H_p = \sum_{m<n}J_{mn}\sigma_z^m\sigma_z^n + \sum_l h_l\sigma_z^l$. The system is initially prepared in the ground state of $H_m= -\sum_i \sigma^i_x$. At the first step ($j=1$), $\alpha(\Delta t)$ is determined from the CD protocols, and a digital step $U_{\rm{cd}}(\Delta t)$ is executed with $\gamma(\Delta t) = 0$. $\gamma(2 \Delta t)$ is computed from Eq.~\eqref{eq:DALCCOeq} by measuring the required expectation values. For $j=2$, we apply the digital block $U_{\rm{cd}}(2 \Delta t)$ and the analog block $U_n(2\Delta t)$ to calculate $\gamma(3 \Delta t)$, continuing similarly for subsequent steps. After $s$ steps, we compute $\braket{H_p}$, resulting in a feedback-based algorithm that utilizes CD protocols and QLC to minimize the energy. A schematic diagram of this algorithm is depicted in Fig.~\ref{fig:schematicalgo}.

We apply DALCCO to several instances of $H_p$ with $J_{mn}$ and $h_{l}$ randomly selected from 
a uniform distribution with support
in $[-1,1]$. In the weak coupling regime, where the condition $J_{mn} \ll h_{l}$ is satisfied, employing local CD protocols yields better results  \cite{li2024quantum,PhysRevResearch.4.043204}. Therefore, this condition is maintained to evaluate the improvements of DALCCO compared to DCQO. Since CD protocols correspond to the digital part of the algorithm,  we restrict the CD operator to the local form $O_1 = \sum_i \sigma_y^i$. The action minimization method is implemented to determine the coefficients $\alpha(t)$ associated with $O_1$ at each step $\Delta t$ (see Supplemental Material).

Regarding the QLC methods, we set $H_n = \sum_j \sigma_y^j +  \sigma_z^j \sigma_x^{j+1}$, where the two-body interaction is the nearest-neighbor cross-resonance, native to transmon-based quantum hardware~\cite{PhysRevLett.107.080502}. Local operators in $H_n$ can be performed digitally as the single-qubit gate errors are negligible compared to two-qubit gates.  
The commutator $i[H_n, H_p]$ results in a 3-local operator, and $i[O_1, H_p]$ results in a 2-local operator whose expectation values are to be measured. The explicit forms of these two operators are given in the Supplemental Material. It is crucial to set $O_1$ such that the commutator results in operators that are as local as possible and with more commuting terms since there will be a measurement overhead based on their exact form. 

We employ a straightforward neighborhood search sub-algorithm to set an optimal value of $f(t)$. Starting with $f(t)=10$, the value is iteratively reduced by a factor of 10 until a monotonically decreasing energy profile is achieved (see Supplemental Material). Hence, we can compute $\alpha(t)$ values at each step from CD protocols and the values of $\gamma(t)$ from the feedback law by using Eq.~\eqref{eq:DALCCOeq}. 

\begin{figure}[t]
    \centering
    \includegraphics[width = \linewidth]{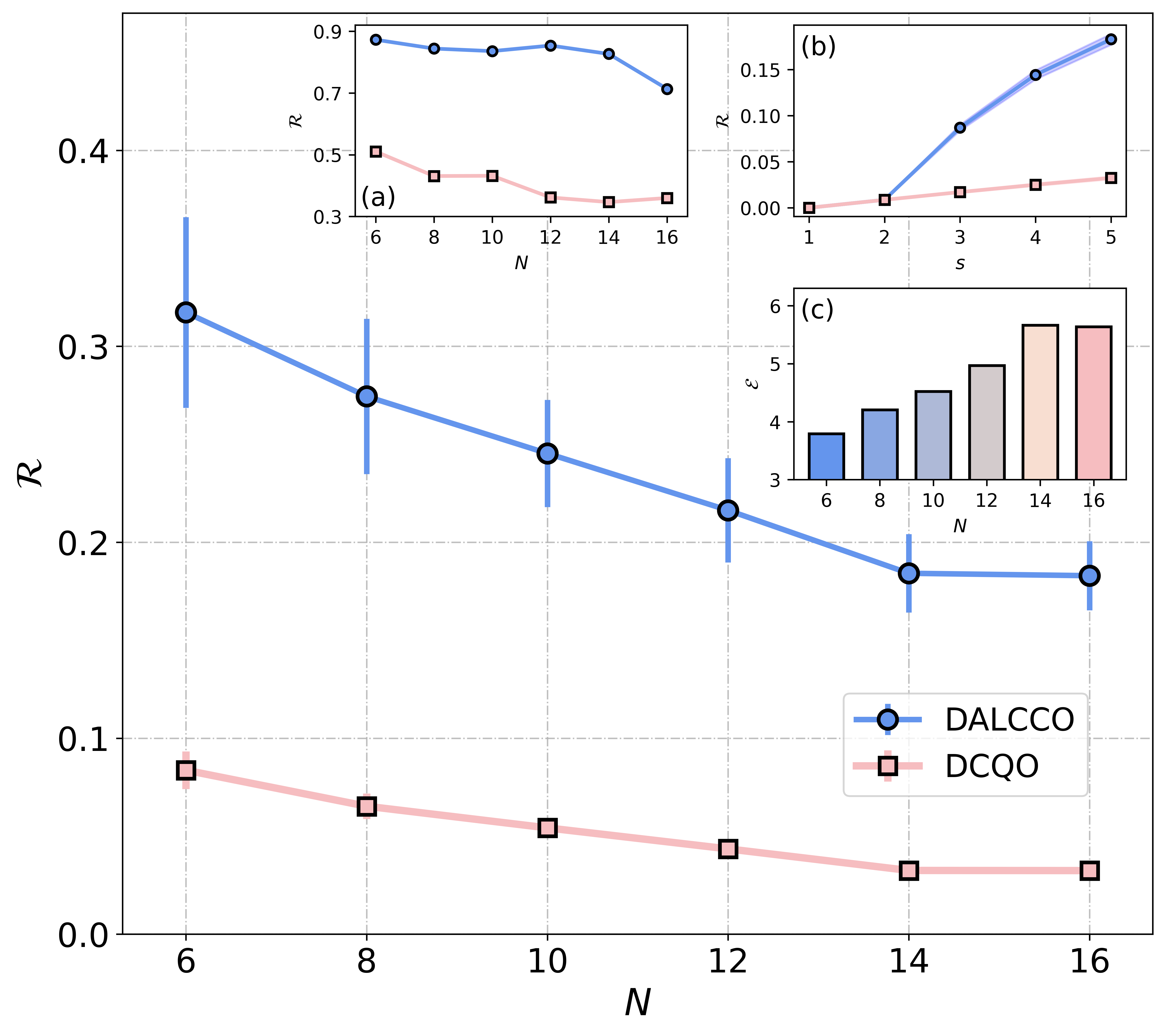}
    \caption{Mean approximation ratio $\mathcal{R}$ as a function of system size $N$ for 500 instances of $H_p$ and $s=5$ steps comparing DALCCO and DCQO. Error bars depict the variance. The inset plot (a) shows the $\mathcal{R}$ values for the best instances, (b) shows  $\mathcal{R}$ as a function of Trotter steps $s$ for the best instance with $N=16$, and (c) shows the enhancement factor $\mathcal{E}$ as a function of $N$ qubits.}
    \label{fig:DALCCOy}
\end{figure}

We define the approximation ratio $\mathcal{R} = \braket{H_p}/E_0$ with $E_0$ being the ground state energy of the system to evaluate the performance. 
Fig.~\ref{fig:DALCCOy} shows the mean $\mathcal{R}$ values of 500 instances of $H_p$, as a function of system size from $N=6$ to $N=16$. The plots show the results of $s=5$ Trotter steps with $\Delta t=0.01$. It is observed that the mean $\mathcal{R}$ values decrease as $N$ increases for both DALCCO and DCQO, but values corresponding to DALCCO are significantly higher. This implies that DALCCO enables us to reach a better approximate solution just by utilizing the native interactions of the quantum hardware. 
To investigate the instance-based performance, the best $\mathcal{R}$ values among all instances of $H_p$ with DALCCO and DCQO are plotted in Fig.~\ref{fig:DALCCOy}(a). For small system sizes, the $\mathcal{R}$ goes as high as $\mathcal{R}\approx 0.9$ and decreases to $\mathcal{R}\approx 0.7$ as the system size increases. Conversely, even in the best instances with DCQO, we get $\mathcal{R}\approx 0.5$ which decreases to $\mathcal{R}\approx 0.3$. This illustrates that DALCCO outperforms DCQO by a large factor, even for the `easy' instances. Additionally, it is apparent from this plot that DALCCO has a higher reachability than DCQO in the sense that it can find solutions that are not accessible by using only CD protocols with the same parameters. The lower $\mathcal{R}$ values for DCQO can be attributed to the low $\Delta t$ values and the small number of Trotter steps. 

In Fig.~\ref{fig:DALCCOy}(b), the mean $\mathcal{R}$ values as a function of Trotter step $s$ for $N=16$ qubits are shown. At $s=1$, the $\mathcal{R}$ values are identical since $\gamma(\Delta t) = 0$ initially. From this point, the $\mathcal{R}$ values increase with each Trotter step, as anticipated. However, it is observed that the slope decreases with increasing $s$. Consequently, at large $s$ values, a plateau-like behavior might be observed. Nonetheless, an enhancement is still evident for all the instances investigated.

We introduce a metric called enhancement factor $\mathcal{E}= \braket{H_p}_{\rm{DALCCO}}/\braket{H_p}_{\rm{DCQO}}$ at $s=5$ to make predictions about the scalability and to quantify the improvements. In Fig.~\ref{fig:DALCCOy}(c), $\mathcal{E}$ as a function of increasing $N$ values is depicted. For $N=6$ we get $\mathcal{E}\approx 4$ which increases to $\mathcal{E}\approx 6$ for $N=16$. This behavior suggests that the decrease rate of $\mathcal{R}$ values is slower for DALCCO than DCQO for increasing $N$. This demonstrates that DALCCO can amplify the $\mathcal{R}$ values up to 5 times on average, and the increase shows signs of potential scalability of the algorithm.

The DALCCO algorithm has many prominent advantages. As previously stated, DALCCO is a digital-analog algorithm, so the two-qubit gate error is negligible in principle. Moreover,  DALCCO bypasses the requirement for classical optimization to find gradients, which is a resource-intensive task and can lead to trainability issues like barren plateaus~\cite{barrenplateau}. Certainly, there can be an equivalent scenario in QLC methods where  $\frac{d}{dt} \braket{H_p}\approx 0$. Under these circumstances, the DALCCO will be dominated by CD protocols and still can find approximate solutions better than DCQO. Although the QLC protocol alone may occasionally be stuck on a zero-gradient energy scale, the combined protocol consistently ensures a reduction in energy at each time step.

One of the most important factors that make our algorithm unique from other CD algorithms is that the choice of $H_n$ is entirely arbitrary except for the condition $ \braket{i[H_n, H_p]} \neq 0$. This opens up numerous possibilities for making the algorithm hardware-viable while keeping it problem-inspired. Furthermore, given that DALCCO combines two quantum control methods, it can find alternative paths to the solution that may not be attained by applying any of the control methods individually. In addition, DALCCO stabilizes the impulse-regime DCQO because QLC can circumvent the non-adiabatic transitions that will occur with the short-time approximation by shifting the state to the lower energy throughout the evolution. 

Despite these advantages, several challenges must be addressed. For example, since this is a stepwise digital-analog algorithm~\cite{PhysRevA.101.022305}, we have to switch the native interactions on and off at each step, which cannot be performed with perfect control. This switching can never be an exact step function, and the system might take some time to stabilize. Also, $\Delta t $ must be sufficiently low to satisfy the QLC condition. This could potentially create a bottleneck since, in some cases, DCQO can provide better solutions with a fixed number of steps. This issue could be addressed by an informed choice of $f(t)$ values. Finding optimal $f(t)$ values generally poses a challenge because it requires evaluating two expectation values at each step to determine the gradient. Therefore, the number of measurements should be taken into account before performing the algorithm. The performance of DALCCO with varying $f(t)$ values is given in Supplemental Material. Apart from this, since this is a feedback-based algorithm, the circuit depth increases with the number of steps. This is an issue for high-depth circuits, but since we are within the impulse regime, DALCCO can still be considered a near-term algorithm. Moreover, the measurements required will always be fewer than those in variational quantum algorithms. These algorithms require many iterations for convergence with multiple measurements at each step to compute gradients~\cite{Wierichs2022generalparameter}. 
\begin{figure}[t]
    \centering
    \includegraphics[width = \linewidth]{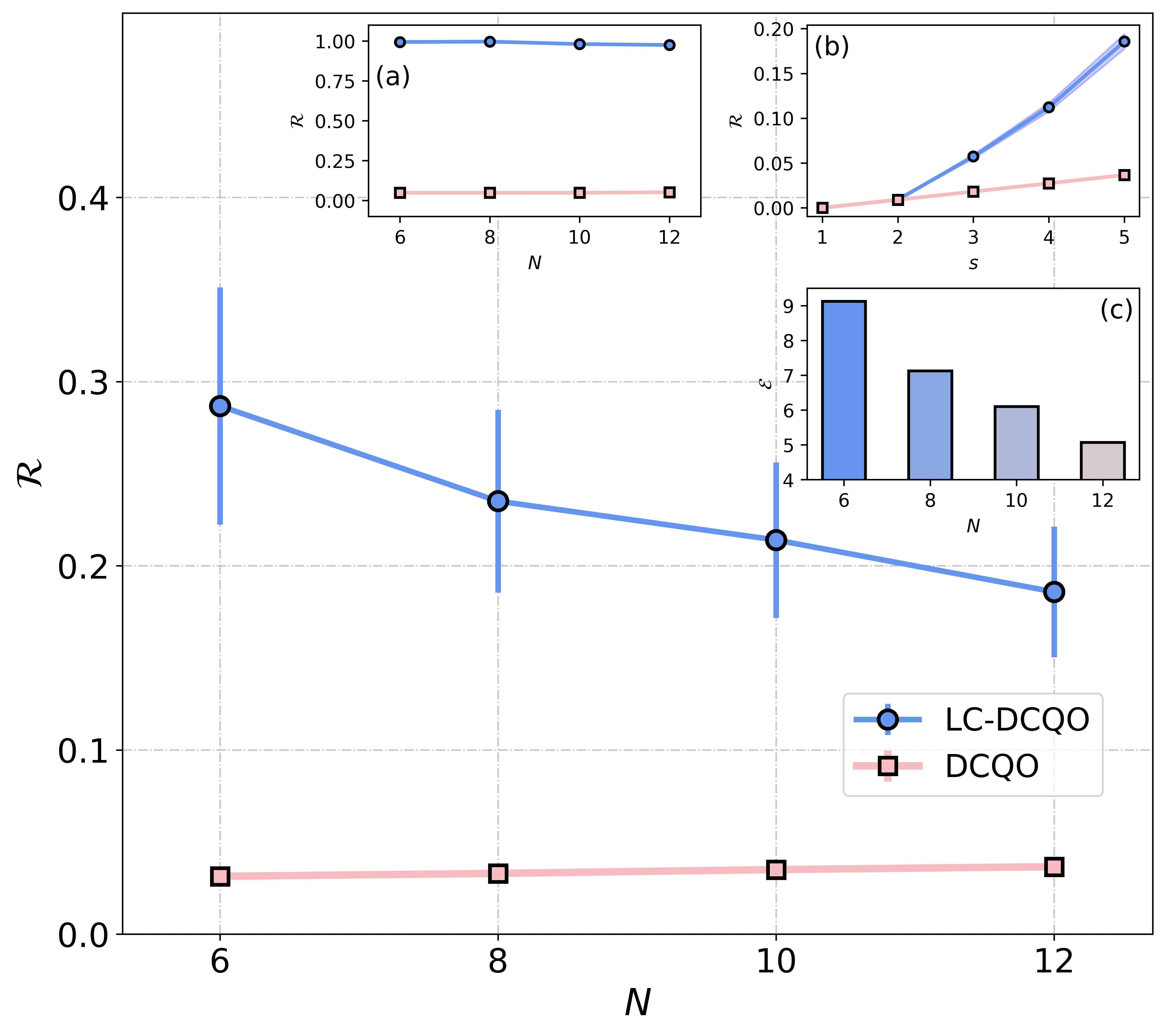}
    \caption{Mean approximation ratio $\mathcal{R}$ as a function of system sizes from $N=6$ to $N=12$ with $s=5$ steps comparing LC-DCQO and DCQO. Inset plots (a) show the $\mathcal{R}$ values for the best instance. (b) shows  $\mathcal{R}$ as a function of Trotter steps $s$ for the best instance for $N=12$, and (c) shows the enhancement factor $\mathcal{E}$ as a function of system size. Error bars show the variance of the 500 instances of $H_p$ considered. }
    \label{fig:DALCCOkry}
\end{figure}

\textit{Digital variant.---} This strategy of combining CD protocols with QLC methods can be seamlessly extended to a purely digital algorithm as well. To differentiate from DALCCO, this algorithm is referred to as Lyapunov-controlled DCQO (LC-DCQO). Unlike DALCCO, in LC-DCQO, there is the liberty to choose non-local AGPs. To this end, we allow up to 2-local terms in the AGPs. Since two-local operators were already utilized in $O_1$, we restrict ourselves to $ H_n=\sum_i \sigma_y^i $. This will maintain the commutator's simplicity so the measurement overhead does not become an issue. The performance of LC-DQCO is compared with DCQO for 500 instances of $H_p$. Here, $J_{mn} \approx h_{l}$, $\Delta t = 0.01$, and $s=5$.

To find the coefficients $\alpha(t)$, we utilize truncated Krylov subspace methods~\cite{bhattacharjee2023lanczos,PhysRevX.14.011032} where the coefficients are found using the Lanczos algorithm for a Krylov dimension $d_K$. However, we limit the expansion at the dimension $d<d_K$ to get $A_\lambda^k = i \sum_k \alpha_k O_{2k-1}$, where $k = \lfloor(d +1)/2\rfloor$. From this expansion, only the first element $A_\lambda^1(t) =  i \alpha(t) O_{1}$ is selected. In our case, we fix $d=5$ for all system sizes with $O_1= \sum_i \Tilde h_i\sigma_y^i +  \sum_{i<j}\Tilde{J}_{ij}(\sigma_y^i\sigma_z^j + \sigma_z^i\sigma_y^j)$, where $\Tilde h_i$ and $\Tilde{J}_{ij}$ are derived using Lanczos algorithm.  In the Supplemental Material, the method is described, and it is shown that the choice of $d$ does not affect the impulse-regime DCQO's performance significantly but reduces the computational complexity to find $\alpha(t)$ values.

In Fig.~\ref{fig:DALCCOkry}, the mean $\mathcal{R}$ values for $N=6$ to $N=12$ qubits is plotted. Like DALCCO, LC-DCQO performs better than DCQO for all the instances considered. In Fig.~\ref{fig:DALCCOkry}(a), the $\mathcal{R}$ values for best-performing instances are illustrated. Note that LC-DCQO can reach $\mathcal{R}=1$ while DCQO is near $\mathcal{R}=0$. This occurs when the ground state is trivial, that is, $\ket{\psi_g} = \ket{0}^{\otimes N}$ or $\ket{\psi_g} = \ket{1}^{\otimes N}$. Therefore, we heuristically identify that LC-DCQO will exhibit significantly better performance when the ground state is near the trivial states. On the other hand, the $\mathcal{R}$ values of DCQO remain nearly constant with increasing $N$. 

In contrast with DALCCO, the enhancement factor $\mathcal{E}$ shown in Fig.~\ref{fig:DALCCOkry}(c) is high but decreases as the system size increases. This behavior can be attributed to the selection of local Hamiltonian $H_n$. However, in Fig.~\ref{fig:DALCCOkry}(b), the slope is increasing; this implies that the results will be better for higher steps $s$. This is due to the small number of steps $s=5$ with step size $\Delta t= 0.01$ and the fact that $J_{mn} \approx h_{l}$ regime will include some hard-to-solve instances with small energy gaps. This will increase the adiabatic times and the time required to solve using CD protocols.
Despite the advantages, there are a few challenges to overcome. Firstly, since this is a digital algorithm, the two-qubit gate errors will play a role if $H_n$ is a non-local Hamiltonian. Also, in LC-DCQO, since the commutator ${i[O_1, H_p]}$ is already non-local, we have to take into account the form of  ${i[H_n, H_p]}$ while choosing $H_n$ to control the measurement overhead.

In summary, we have introduced a digital-analog quantum algorithm, DALCCO, and a purely digital algorithm, LC-DCQO, both designed to tackle combinatorial optimization problems and eliminate classical optimization.
They integrate quantum Lyapunov control with counterdiabatic protocols, providing an advantage in approximation ratios compared to impulse-regime counterdiabatic methods. This can be achieved by maintaining the same digital circuit depth and leveraging measurements as a resource. 
Our results show that the proposed algorithms perform considerably better than DCQO techniques, which have shown empirical polynomial improvements over adiabatic quantum algorithms~\cite{cadavid2024bias}. This work has the potential to achieve further digital compression using the hardware's native interactions. In addition, this approach improves local CD driving, which is an active area of research~\cite{li2024quantum,morawetz2024efficient}.

{\it Acknowledgements.} P.C. and K.P. are grateful to Mikel Garcia-de-Andoin for useful comments. This work is supported by EU FET Open Grant  EPIQUS (899368) and the Basque Government through Grant No. IT1470-22, the Spanish Ministry of Economic Affairs and Digital Transformation through the QUANTUM ENIA project call-Quantum Spain project, and the IKUR Strategy of the Basque Government under the collaboration agreement between the Ikerbasque Foundation and the University of the
Basque Country. This project was supported by the Luxembourg National Research Fund (FNR Grant Nos. 17132054 and 16434093). It has also received funding from the QuantERA II Joint Programme and co-funding from the European Union’s Horizon 2020 research and innovation programme.

\bibliography{reference.bib}

\ifarXiv
    

\section*{Supplementary material}

# Supplemental Material: Lyapunov Controlled Counterdiabatic Quantum Optimization

Pranav Chandarana 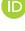,[1, 2, *] Koushik Paul 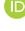,[1, 2, †] Kasturi Ranjan Swain 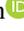,[3] Xi Chen 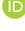,[4, ‡] and Adolfo del Campo 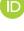[3, 5, §]

[1]*Department of Physical Chemistry, University of the Basque Country UPV/EHU, 48080 Bilbao, Spain*
[2]*EHU Quantum Center, University of the Basque Country UPV/EHU, 48940 Leioa, Spain*
[3]*Department of Physics and Materials Science, University of Luxembourg, L-1511 Luxembourg, Luxembourg*
[4]*Instituto de Ciencia de Materiales de Madrid (CSIC), Cantoblanco, E-28049 Madrid, Spain*
[5]*Donostia International Physics Center, E-20018 San Sebastián, Spain*

## FINDING COEFFICIENTS TO SATISFY LYAPUNOV CONDITION

In this section, we explicitly derive how to obtain the coefficients $\gamma(t)$ that satisfy the Lyapunov condition $\frac{d}{dt}\langle H_p \rangle \leq 0$. To accomplish this task, we start with

$$H_l(t) = A^k(t) + \gamma(t)H_n. \tag{1}$$

Putting $A_\lambda^k = \alpha(t)O_k$, the Lyapunov condition can be expressed as

$$\frac{d}{dt}\langle H_p \rangle = \lambda(t)\alpha(t)\langle i[O_k, H_p]\rangle_t + \gamma(t)\langle i[H_n, H_p]\rangle_t \leq 0. \tag{2}$$

We noticed that for all the instances we consider, if $\langle i[O_k, H_p]\rangle_t > 0$, we have $\alpha(t) < 0$ and vice versa. Hence, Eq. (2) can be rewritten as

$$\frac{d}{dt}\langle H_p \rangle = \lambda(t)\alpha(t)\langle i[O_k, H_p]\rangle_t \left(1 + \frac{\gamma(t)\langle i[H_n, H_p]\rangle_t}{\lambda(t)\alpha(t)\langle i[O_k, H_p]\rangle_t}\right) \leq 0. \tag{3}$$

As $\alpha(t)\langle i[O_k, H_p]\rangle < 0$, one can construct $\gamma(t)$ such that

$$1 + \frac{\gamma(t)\langle i[H_n, H_p]\rangle_t}{\alpha(t)\lambda(t)\langle i[O_k, H_p]\rangle_t} \geq 0. \tag{4}$$

For that we set,

$$\gamma(t) = f(t)\langle i[H_n, H_p]\rangle_t \langle i[\lambda(t)\alpha(t)O_k, H_p]\rangle_t, \tag{5}$$

where $f(t)$ is a Lagrangian multiplier. Putting the value of $\gamma(t)$ back in Eq. (3), we get the condition $1 + f(t)|\langle i[H_n, H_p]\rangle_t|^2 \geq 0$. Hence, we select $f(t)$ as a positive constant such that Eq. (4) is satisfied. Regardless of the sign of $\alpha(t)$, we can always select the appropriate sign of $f(t)$, maintaining the condition.

In Fig 1, we plot the $\gamma(t)$ values for nine different instances for $H_p$ with $N = 16$ qubits obtained implementing DALCCO. We observe that the behavior of $\gamma(t)$ highly depends on the instances chosen and shows the small positive values for most instances. However, in some instances, they go to very high values and can also take negative ones. However, we can always find a positive time corresponding to these evolutions by simple transformations.

## NEIGHBORHOOD SEARCH SUB-ALGORITHM

We implement a neighborhood search sub-algorithm to find an optimal $f(t)$ value. In this sub-algorithm, we start with $f(t) = 10$ and check for a monotonic decrease in the energy for $s$ steps. If it does decrease, we check $[2f(t), 3f(t), 4f(t), 5f(t)]$ and select the $f(t)$ that gives minimum energy. If it does not decrease, we divide $f(t)$ by 10 to check again. We do the same neighborhood search for the new $f(t)$ as well. This will give us the best $f(t)$ values within a certain neighborhood. Note that this algorithm is heuristically designed, taking into account the general behavior of $f(t)$ values for all the instances we consider. It should be noted that we have to calculate the expectation values of two operators $i[H_n, H_p]$ and $i[O_1, H_p]$ for each step. Therefore, designing these algorithms also should take into account tradeoffs between the measurement overhead and performance.

To realize how the $\langle H_p \rangle$ converges with varying $f(t)$ values, we consider the best instance of $N = 16$ system size using DALCCO. In Fig. 2, we plot $\langle H_p \rangle$ as a function of steps $s$ with varying $f(t)$ values. We observe that for different values of $f(t)$, the energy does not decrease monotonically except for two. Among these two the algorithm will select the $f(t)$ with lower final energy. For the first iteration, it is the same as $\gamma(0\Delta t) = 0$.

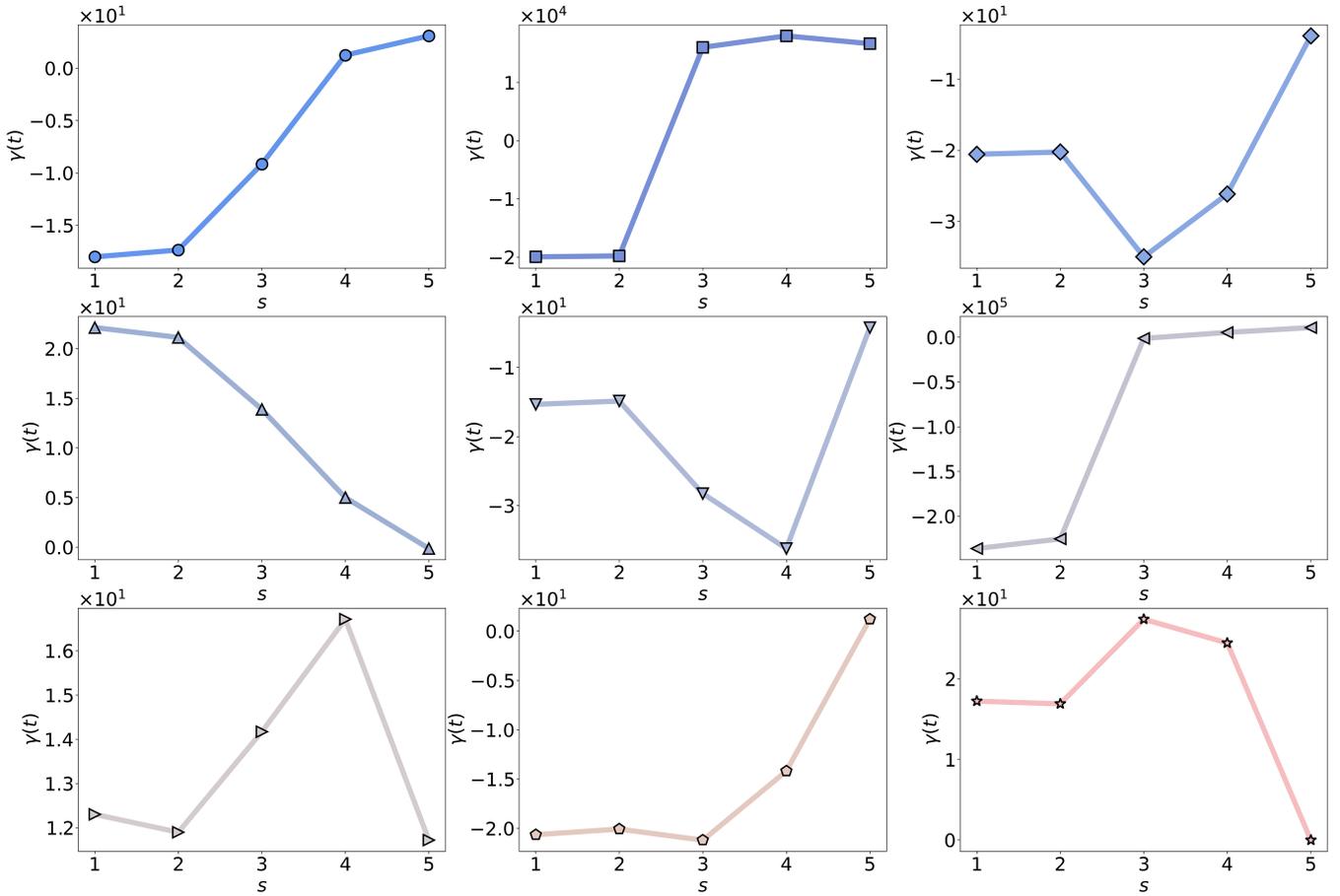

FIG. 1. Lyapunov coefficients $\gamma(t)$ for different instances of $H_p$ as a function of Trotter step $p$ implementing DALCCO with $N = 16$ qubits.

---

**Algorithm 1** Neighborhood search algorithm to find best $f(t)$ values

---

1: **Input:** Initial function value $f(t) = 10$, number of steps $s$
2: **Output:** Optimized function value $f(t)$
3: **for** $i = 1$ to $s$ **do**
4:     **if** Energy decreases monotonically **then**
5:         Check $[2f(t), 3f(t), 4f(t), 5f(t)]$
6:         Select the $f(t)$ that gives the minimum energy
7:     **else**
8:         $f(t) \leftarrow \frac{f(t)}{10}$
9:     **end if**
10:     Perform the same neighborhood search for the new $f(t)$
11: **end for**

---

## ACTION MINIMIZATION METHOD

The AGP operator $A_\lambda^k$ in Eq. 1 of the main text satisfies,

$$[H_a, i\partial_\lambda H_a + [H_a, A_\lambda^k]] = 0. \tag{6}$$

As shown in the main text, the approximate AGP can be written as $A_\lambda^k = i\sum_k \alpha_k O_{2k-1}$, which is determined by a set of coefficients $\{\alpha_k\}$. For finite $k$, $\{\alpha_k\}$ can be obtained variationally by minimizing the action,

$$S_k = Tr[G_k{}^2] \quad \text{with} \quad G_k = \partial_\lambda H_a - i[H_a, A_\lambda^k]. \tag{7}$$

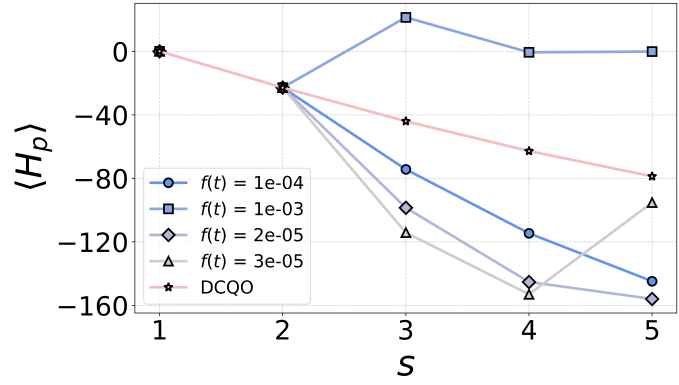

FIG. 2. Energies $\langle H_p \rangle$ as a function of steps $p$ with various $f(t)$ values for $N = 16$ system size

Performing this with $O_1 = \sum_i \sigma_y^i$, we get

$$\alpha(t) = \frac{1}{2} \frac{\sum_{i=1}^{N} h_i}{[1 - \lambda(t)]^2 N + \lambda^2(t) \sum_i h_i^2 + 2\lambda^2(t) \sum_{i<j} J_{ij}^2}. \tag{8}$$

## TRUNCATED KRYLOV SUBSPACE METHOD

This section describes how to calculate approximate gauge potential using truncated Krylov subspace methods. According to Baker-Campbell-Hausdorff formula, the time evolution of an operator $O(t)$ can be expressed as

$$\begin{aligned} O(t) &= e^{iHt/\hbar} O e^{-iHt/\hbar} \\ &= O + \frac{it}{\hbar}[H, O] + \frac{(it)^2}{2! \hbar^2}[H, [H, O]] + \frac{(it)^3}{3! \hbar^3}[H, [H, [H, O]]] + \cdots \\ &= O + \frac{it}{\hbar}\mathcal{L}(O) + \frac{(it)^2}{2! \hbar^2}\mathcal{L}^2(O) + \frac{(it)^3}{3! \hbar^3}\mathcal{L}^3(O) + \cdots, \end{aligned} \tag{9}$$

where $\mathcal{L}$ shows the Louivillian super-operator. The terms in the expansion can be considered as basis elements that span the entire subspace. However, these basis states may not be orthonormal, and one can use the Gram-Schmidt orthonormalization procedure or Lanczos algorithm to generate the normalized Krylov basis. We start by defining an inner product between two arbitrary operators $A$ and $B$ as,

$$(A, B) = \frac{1}{2} \text{Tr}\left[\rho(H)(A^\dagger B + BA^\dagger)\right], \tag{10}$$

where $\rho(H)$ is a positive-definite Hermitian operator. The Lanczos coefficients $\{b_n\}$ and the Krylov operators $\{O_n\}$ are obtained numerically as described in Algorithm 2, where we truncate the expansion at an arbitrary odd Krylov dimension $d$ before $b_n$ hits zero. Substituting $A_\lambda^k = i \sum_k \alpha_k O_{2k-1}$ into Eq. (6) yields the matrix equation,

$$\begin{bmatrix} b_1^2 + b_2^2 & b_2 b_3 & 0 & 0 & \cdots & 0 \\ b_2 b_3 & b_3^2 + b_4^2 & b_4 b_5 & 0 & \cdots & 0 \\ 0 & b_4 b_5 & b_5^2 + b_6^2 & b_6 b_7 & \cdots & 0 \\ \vdots & \vdots & \vdots & \vdots & \ddots & \vdots \\ 0 & 0 & 0 & \cdots & b_{d-3} b_{d-2} & b_{d-2}^2 + b_{d-1}^2 \end{bmatrix} \begin{bmatrix} \alpha_1 \\ \alpha_2 \\ \alpha_3 \\ \vdots \\ \alpha_{d_A} \end{bmatrix} = \begin{bmatrix} -b_0 b_1 \\ 0 \\ 0 \\ \vdots \\ 0 \end{bmatrix}. \tag{11}$$

Next, using $\{b_n\}$ the AGP operators can be obtained by inverting the matrix of dimension $d_A \times d_A$, where $d_A = \lfloor d/2 \rfloor$. In this work, we restrict ourselves to $\rho(H) = 1$ and only to the first order term $A_\lambda^1 = i\alpha O_1$, since it will be difficult to implement the higher order terms on a quantum computer.

---

**Algorithm 2** Finding the AGP terms by Krylov subspace construction

---

$B_0 = \partial_\lambda H$
$b_0 = \sqrt{(B_0, B_0)}$
$O_0 = B_0/b_0$
$B_1 = \mathcal{L}(O_0)$
$b_1 = \sqrt{(B_1, B_1)}$
$O_1 = B_1/b_1$
**for** $n = 2$ to $d$ **do**
  $B_n = \mathcal{L}(O_{n-1}) - b_{n-1}O_{n-2}$
  $b_n = \sqrt{(B_n, B_n)}$
  $O_n = B_n/b_n$
**end for**
Using $\{b_n\}$ solve the matrix Eq. (11) to obtain $\{\alpha_k\}$
$A_\lambda^1 = i\alpha O_1$

---

As shown in Fig. 3, the performance of the LC-DCQO algorithm does not depend significantly on the truncated Krylov dimension $d$. Although the maximum of $|\alpha(t)|$ coefficients shows an increment of $\sim 42\%$ for $d = 21$ as compared to $d = 5$ for system size $N = 8$. The fidelity for $d = 21$ with respect to the ground state of $H_p$ only increases by $\sim 2\%$ compared to $d = 5$ at the end of $s = 20$ Trotter steps. Therefore, we perform all our calculations for $d = 5$ in the main text as choosing a lower $d$ reduces the computational complexity to invert the matrix in Eq. (11). Fig. 4 illustrates the average $|\alpha(t)|$ coefficients obtained for the same 500 instances used in the main text. The comparison is shown for the action minimization and truncated Krylov subspace methods. The CD coefficients obtained from the latter are larger because of the normalization coefficients $b_0$, $b_1$ with Krylov subspace methods.

### Decomposing $O_1$

This subsection provides details about the calculation of the Krylov operator $O_1$ and decomposing it in terms of Pauli operators. Following the Krylov construction in Algorithm. 2, we have $O_1 = \frac{1}{b_1}\mathcal{L}(\partial_\lambda H/b_0)$. Taking the adiabatic Hamiltonian (as given in Eq. 1 of main text), we get,

$$
\begin{aligned}
O_1 = \frac{1}{b_0 b_1}\mathcal{L}(\partial_\lambda H_a(t)) &= \frac{1}{b_0 b_1}[(1 - \lambda(t))H_m + \lambda(t)H_p, -H_m + H_p] \\
&= \frac{1}{b_0 b_1}\left((1 - \lambda(t))[H_m, H_p] + \lambda(t)[H_m, H_p]\right) \\
&= \frac{1}{b_0 b_1}[H_m, H_p] \\
&= \frac{1}{b_0 b_1}\left[-\sum_i \sigma_i^x, \sum_{m<n} J_{mn}\sigma_z^m\sigma_z^n + \sum_l h_l\sigma_z^l\right] \\
&= \frac{1}{b_0 b_1}\left(2i\sum_{i<j} J_{ij}(\sigma_y^i\sigma_z^j + \sigma_z^i\sigma_y^j) + 2i\sum_i h_i\sigma_y^i\right) \\
&= i\sum_{i<j} \tilde{J}_{ij}(\sigma_y^i\sigma_z^j + \sigma_z^i\sigma_y^j) + i\sum_i \tilde{h}_i\sigma_y^i,
\end{aligned}
\tag{12}
$$

where $\tilde{J}_{ij} = 2J_{ij}/b_0 b_1$ and $\tilde{h}_i = 2h_i/b_0 b_1$.

### COMMUTATOR CALCULATION

In this section, we calculate expectation value of $i\left[H_n, H_p\right]$ and $i\left[O_1, H_p\right]$ as required by DALCCO. Let us begin with $i\left[H_n, H_p\right]$ given by

$$
i\left[H_n, H_p\right] = \sum_{ij}\sum_{mn}\underbrace{\left[Z_i X_j, J_{mn}Z_m Z_n\right]}_{\textbf{(a)}} + \sum_{ij}\sum_k\underbrace{\left[Z_i X_j, h_k Z_k\right]}_{\textbf{(b)}} + \sum_{ij}\sum_k\underbrace{\left[Y_k, J_{ij}Z_i Z_j\right]}_{\textbf{(c)}} + \sum_{ij}\sum_k\underbrace{\left[Y_i, h_k Z_k\right]}_{\textbf{(d)}}.
\tag{13}
$$

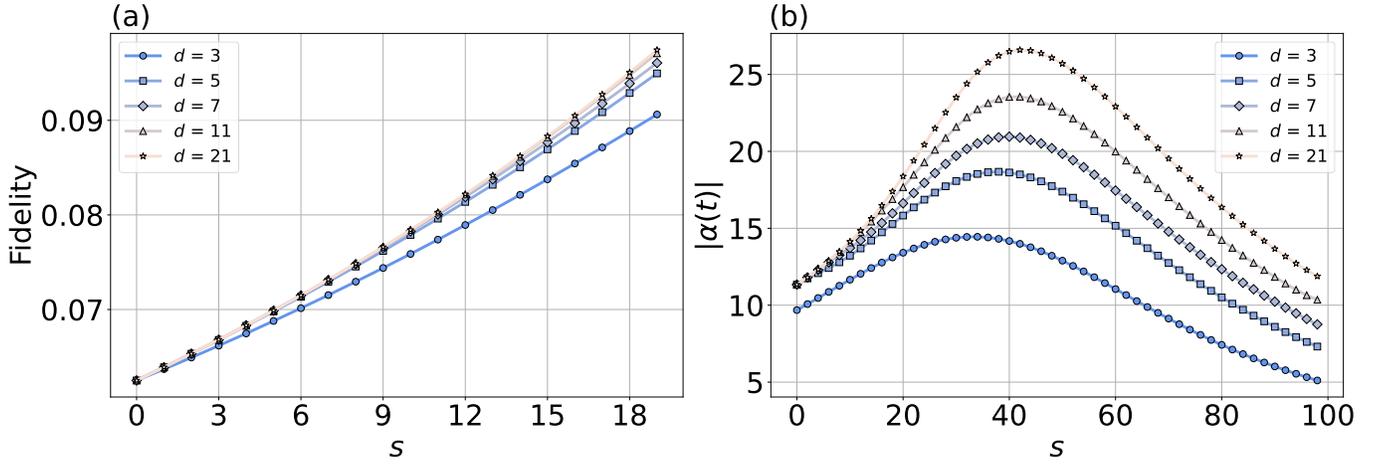

FIG. 3. LC-DCQO simulations of system size $N = 8$ at truncated Krylov dimension $d = 3, 5, 7, 11, 21$ averaged over 20 instances. (a) The fidelity with respect to the ground state of $H_p$ as a function of $s = 20$ Trotter steps. (b) $|\alpha(t)|$ coefficients as a function of $s = 100$ Trotter steps.

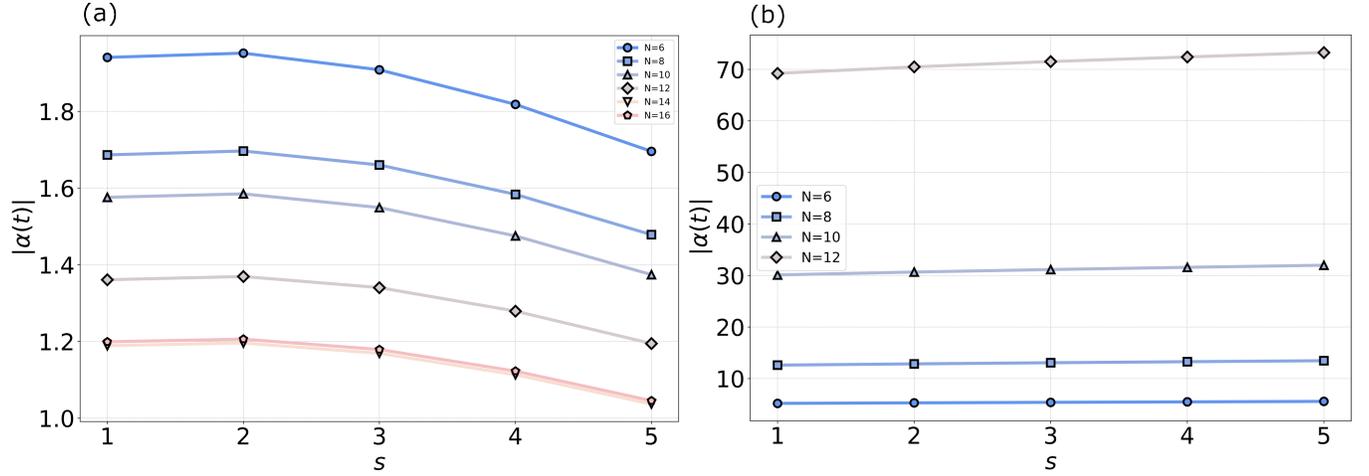

FIG. 4. Average CD coefficients $|\alpha(t)|$ as a function of $s = 5$ Trotter steps using (a) DALCCO with action minimization method and (b) LC-DCQO with truncated Krylov method. Plots show an average of 500 instances considered in the main text for different system sizes.

Nest, we calculate the commutator expansion term by term,

$$
\begin{aligned}
\textbf{(a)} \quad \sum_{ij} \sum_{mn} \Big[ Z_i X_j, J_{mn} Z_m Z_n \Big] &= \sum_{ij} \sum_{mn} Z_i \Big[ X_j, J_{mn} Z_m Z_n \Big] = \sum_{ij} \sum_{mn} Z_i \Big( 2\delta_{jm} J_{mn} Y_j Z_n + 2\delta_{jn} J_{mn} Z_m Y_j \Big) \\
&= \sum_{ij} \sum_{mn} +2 J_{mn} \delta_{jm} Z_i Y_j Z_n + 2 J_{mn} \delta_{jn} Z_i Z_m Y_j \\
&= \sum_{ij} \sum_{mn} +2 J_{mn} \delta_{jm} Z_i Y_j Z_n + 2 J_{mn} \delta_{jn} \delta_{im} Y_j + 2 J_{mn} \delta_{jn} (1 - \delta_{im}) Z_i Z_m Y_j,
\end{aligned}
$$

$$
\textbf{(b)} \quad \sum_{ij} \sum_{k} \Big[ Z_i X_j, h_k Z_k \Big] = \sum_{ij} \sum_{k} Z_i \Big[ X_j, h_k Z_k \Big] = \sum_{ij} \sum_{k} +2\delta_{jk} h_k Z_i Y_j,
$$

$$
\textbf{(c)} \quad \sum_{ij} \sum_{k} \Big[ Y_k, J_{ij} Z_i Z_j \Big] = \sum_{ij} \sum_{k} -2\delta_{ki} J_{ij} X_k Z_j - 2\delta_{kj} J_{ij} Z_i X_k,
$$

$$
\textbf{(d)} \quad \sum_{ij} \sum_{k} \Big[ Y_i, h_k Z_k \Big] = \sum_{ij} \sum_{k} -2\delta_{ik} h_k X_i.
$$

Finally, $i\left[O_1, H_p\right]$ can be calculated as

$$i\left[O_1, H_p\right] = \sum_{ij}\sum_k \left[Y_k, J_{ij}Z_iZ_j\right] + \left[Y_i, h_kZ_k\right] = \sum_{ij}\sum_k -2\delta_{ki}J_{ij}X_kZ_j - 2\delta_{kj}J_{ij}Z_iX_k - 2\delta_{ik}h_kX_i. \tag{14}$$

* pranav.chandarana@gmail.com
† koushikpal09@gmail.com
‡ xi.chen@csic.es
§ adolfo.delcampo@uni.lu


\fi
\end{document}